\documentclass[aps,prb,twocolumn,superscriptaddress,showpacs]{revtex4-1}
\usepackage{graphicx}
\usepackage{amsmath,amssymb}

\begin{document}

\title{Bond Randomness Induced Magnon Decoherence in a Spin-1/2 Ladder Compound}
\author{B.~N\'afr\'adi}
\affiliation{Max-Planck-Institut f\"ur Festk\"orperforschung, Heisenbergstra\ss e 1, D-70569 Stuttgart, Germany}
\author{T.~Keller}
\affiliation{Max-Planck-Institut f\"ur Festk\"orperforschung, Heisenbergstra\ss e 1, D-70569 Stuttgart, Germany}
\author{H.~Manaka}
\affiliation{Graduate School of Science and Engineering, Kagoshima University, Kagoshima 890-0065,Japan}
\author{U.~Stuhr}
\affiliation{Laboratory for Neutron Scattering, Paul Scherrer Institute, CH-5232 Villigen PSI, Switzerland}
\author{A.~Zheludev}
\affiliation{Neutron Scattering and Magnetism Group, Laboratorium f\"ur Festk\"orperphysik, ETH Z\"urich, CH-8093, Switzerland}
\author{B.~Keimer}
\affiliation{Max-Planck-Institut f\"ur Festk\"orperforschung, Heisenbergstra\ss e 1, D-70569 Stuttgart, Germany}

\date{\today}

\begin{abstract}

We have used a combination of neutron resonant spin-echo and triple-axis spectroscopies to determine the energy and linewidth of the magnon resonance in IPA-Cu(Cl$_{0.95}$Br$_{0.05}$)$_3$, a model spin-1/2 ladder antiferromagnet where Br substitution induces bond randomness. 
We find that the bond defects induce a blueshift, $\delta \Delta$, and broadening, $\delta \Gamma$, of the magnon gap excitation compared to the pure compound. 
At temperatures exceeding the energy scale of the inter-ladder exchange interactions, $\delta \Delta$ and $\delta \Gamma$ are temperature independent within the experimental error, in agreement with Matthiessen's rule according to which magnon-defect scattering yields a temperature independent contribution to the magnon mean free path. 
Upon cooling, $\delta \Delta$ and $\delta \Gamma$ become temperature dependent and saturate at values lower than those observed at higher temperature, consistent with the crossover from one-dimensional to two-dimensional spin correlations with decreasing temperature previously observed in pure IPA-CuCl$_3$. 
These results indicate limitations in the applicability of Matthiessen's rule for magnon scattering in low-dimensional magnets.

\end{abstract}


\maketitle

%
One-dimensional (1D) spin liquids, such as spin-1 chains or spin-1/2 ladders with antiferromagnetic interactions, have emerged as important model systems for the study of long-range quantum coherence \cite{Altshuler2006}.
These systems have collective spin-singlet ground states. 
Their magnetic excitations are mobile, triply degenerate ``magnon'' quasiparticles, with a gap $\Delta_0$ even for temperature $T \rightarrow 0$. 
In an ideally clean system the magnon mean free path is infinitely long at $T=0$. 
Heat transport measurements have indeed reported magnon mean-free path as long as several hundred unit cells \cite{Hess_PRB_2001}. 
At nonzero temperatures, theoretical work based on the non-linear sigma model (NL$\sigma$M) predicts that the temperature dependent coherence length, and thus observables such as the energies, $\Delta$, and linewidths, $\Gamma$, of the magnons depend solely on $T/\Delta_0$ \cite{Senechal1993,Sachdev1997,Zheludev2008}. 
The NL$\sigma$M yields universal prediction for $\Delta (T)$ and $\Gamma (T)$ in 1D spin liquids, which have been verified for spin-1 chains \cite{Xu2007} and spin-1/2 ladders \cite{Zheludev2008,Nafradi2011}.

Here we experimentally address the influence of bond disorder on the spin dynamics of a spin-1/2 ladder compound, an issue that has recently attracted considerable attention in view of the possible realization of a novel ``Bose glass'' phase in sufficiently intense magnetic fields. \cite{Hong2010,Crepin2011,Carrasquilla2011,Hong2010b}
In zero field, quantum Monte Carlo calculations have shown that bond disorder decreases the magnon coherence length, \cite{Trinh_PRB_2012,Greven_PRL_1998} and
it was shown experimentally that in disordered 1D spin-1 chains at $T = 0 $ chemical defects limit the coherence length in a manner similar to thermally excited mobile magnons at $T \neq 0$ \cite{Xu2007}.
Accordingly, thermal conductivity data have been analyzed in terms of a simple superposition of $T$-dependent and $T$-independent scattering rates (``Matthiesen's rule'') \cite{Hess_PRL_2004,Hess_PRB_2006}. 
A simple yet effective picture is that magnons are confined into finite 1D potential ``boxes'', defined by both static and dynamic defects.
This picture is limited to one dimension, \cite{Sachdev1997,Zheludev2008} and is less rigorously justified in the presence of inter-chain or inter-ladder interactions which are present in real materials \cite{Nafradi2011}. 
Nonetheless, thermal conductivity data on compounds with 2D spin systems were also analyzed in terms of an analogous Matthiesen rule expression \cite{Hess_PRL_2004}. 
The goal of our spectroscopic experiments is to test the applicability of this simple rule for the energy, $\Delta$, and width, $\Gamma$, of the magnon resonance in a spin-1/2 ladder system.

Bromine substituted IPA-Cu(Cl$_{1-x}$Br$_x$)$_3$ is an excellent model system to experimentally study the effect of bond disorder on the energies and linewidths of the magnon resonance in spin-1/2 ladders with antiferromagnetic interactions.
For $x < 13\% $, this compound crystallizes in the triclinic space group $P \bar{1}$ \cite{Manaka2001e,Manaka1997c}.
We index the momentum-space coordinates $\mathbf{q} = (h,k,l)$ in the corresponding reciprocal lattice units (r.l.u.).
In the \textit{a-c} plane the nonmagnetic IPA \footnote{Here IPA stands for isopropyl ammonium, which in our case was fully deuterated: (CD$_3$)$_2$CDND$_3$.} molecules form layers so that the magnetic interactions along \textit{b} are negligible \cite{Masuda2006,Hong2010}.
The Cu$^{2+}$ ions carry a spin $S=1/2$ and form ladders along the \textit{a} axis.
Exchange interactions within these Cu$^{2+}$ ions are mediated by Cl di-bridges.
Substituting Cl with Br results in a slight change of the bond angles but a substantial change in the interaction strength, as discussed in detail elsewhere \cite{Hida2003,Manaka2008}.
This modification does not directly involve the spin-carrying Cu$^{2+}$ ions and does not add additional local $ S=1/2 $ which would dominate the low temperature magnetic properties.
Thus IPA-Cu(Cl$_{0.95}$Br$_{0.05}$)$_3$ remains in a gapped quantum spin liquid state with $ \Delta_0=1.24$~meV at $\mathbf{q} = (0.5,0,0)$ at $ T=1.5$~K which is higher than in the pure IPA-CuCl$_3$ system \cite{Hong2010}.
Just like in pure IPA-CuCl$_3$, the primary dispersion of the magnons along \textit{h} is well reproduced by a spin-ladder model with a ferromagnetic coupling $J_1 = -2.3$~meV along the rungs and antiferromagnetic couplings $J_2 = 1.2$~meV and $ J_3=2.9 $~meV along the legs and diagonals of the ladders of the ladders, respectively \cite{Fischer2011}.
Parallel ladders are coupled by the ferromagnetic interladder interaction $ J_4 = -0.3 $~meV.
The quasiparticle spectrum terminates at a critical wave vector $ h_c \sim 0.2 $ \cite{Hong2010}.

%
The sample used for the experiments was an assembly of five fully deuterated single crystals of IPA-Cu(Cl$_{0.95}$Br$_{0.05}$)$_3$ of total volume $\sim 0.8$~g used also in Ref.~\onlinecite{Hong2010}.
The preparation method is described elsewhere \cite{Manaka2001e}.
The neutron resonant spin-echo (NRSE-TAS) experiments were performed on the spectrometer TRISP at the research reactor FRM-II in Garching, Germany \cite{Keller2002}.
The incident neutron beam was spin-polarized by a supermirror guide.
The (002) reflection of pyrolytic graphite was used to select the energies of incident and scattered neutrons.
A velocity selector was used to cut out higher-order contamination of the incident beam.
The polarization of the scattered beam with a fixed final energy $E_f=6.7$ meV was measured by a transmission polarizer in front of the detector.
The energy resolution of these measurements was about $ 1~\mu$eV \cite{Keller2002,Nafradi2011}.
The triple-axis spectroscopie (TAS) measurements were carried out at the TASP cold neutron spectrometer at PSI in Villigen, Switzerland, with  neutrons of fixed final energy $E_f=5$ meV.
The beam optics included vertically focused pyrolytic graphite monochromators and horizontally focused analyzers, as well as Be filters positioned after the sample.
The energy resolution of this setup was about 0.16 meV.

%
\begin{figure}
    \includegraphics[width=7.9cm]{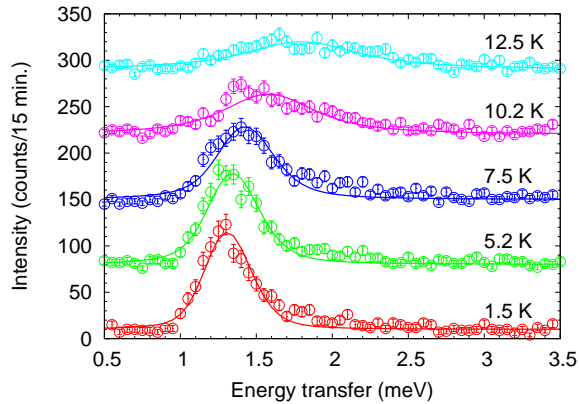}
   \caption{(Color online) Typical constant-q scans measured on a cold-TAS spectrometer in IPA-Cu(Cl$_{0.95}$Br$_{0.05}$)$_3$ at $ q=(1.5,0,0)$ at different temperatures. The solid lines are the results of fits to Lorentzians convoluted with the resolution function of the instrument. Scans at different temperatures are shifted vertically for clarity. \label{fig:1}}
\end{figure}

\begin{figure}
    \includegraphics[width=8.5cm]{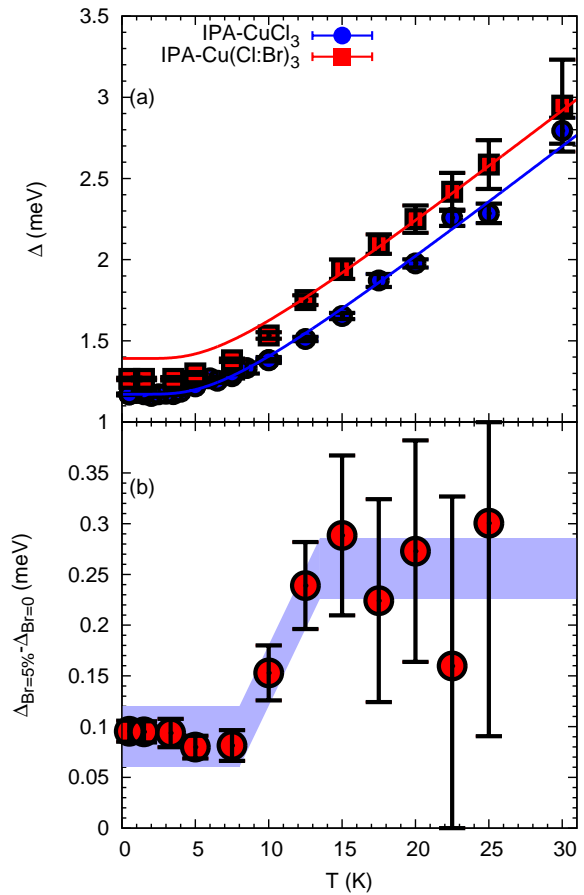}
   \caption{(Color online) (a) Temperature dependence of the magnon energy gap $\Delta$ for IPA-Cu(Cl$_{0.95}$Br$_{0.05}$)$_3$ (red points). Blue points are for IPA-CuCl$_3$ from Refs.~\onlinecite{Nafradi2011,Zheludev2008}. The blue line is the result of a calculation based on a NL$\sigma$M with $\Delta_0 = 1.44$~meV and interladder interaction $J_\perp / J =0.1$, following Ref.~\onlinecite{Senechal1993a}. The red line is shifted by $ 0.25 $~meV relative to the blue one. (b) Temperature dependence of the bond randomness induced magnon blueshift of the 5\% Br substituted compound. The broad shaded line is a guide to the eyes.\label{fig:gap}}
\end{figure}

\begin{figure}
    \includegraphics[width=8.5cm]{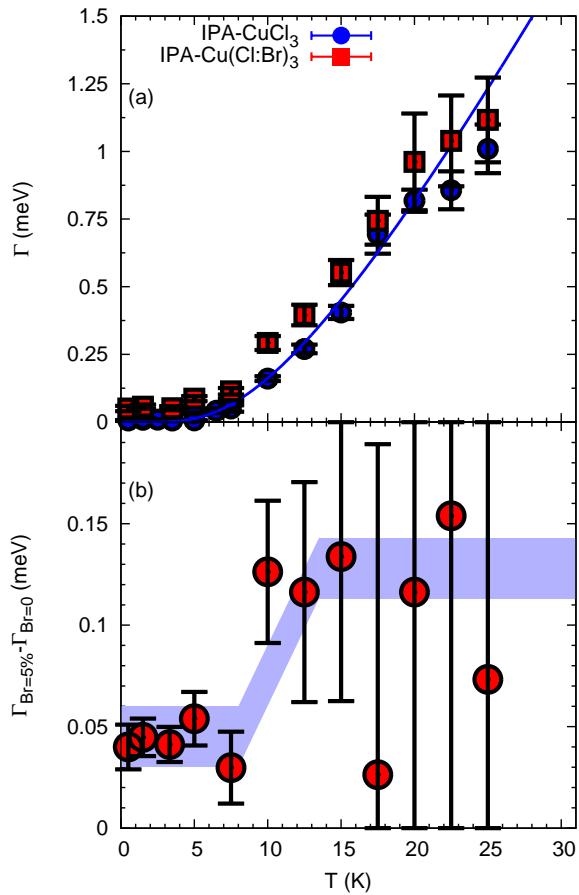}
    \caption{(Color online) (a) Temperature dependence of the magnon line width $\Gamma$. Blue points are from Refs.~\onlinecite{Nafradi2011,Zheludev2008}. Red points are the result of the present study. The line represents NL$\sigma$M calculations \cite{Damle1998} with $\Delta_0=1.44$~meV. (b) Temperature dependence of the magnon decoherence induced by bond randomness of the 5\% Br substituted compound. The broad shaded line is a guide to the eyes.\label{fig:3}}
\end{figure}

NRSE-TAS data at the bottom of the magnon dispersion relation at momentum $ \mathbf{q} = (0.5,0,0) $ and temperature $ T = 0.5 $~K for IPA-Cu(Cl$_{0.95}$Br$_{0.05}$)$_3$ revealed a single magnon resonance with $ \Gamma_{5\%~\mathrm{Br}} (0.5~\mathrm{K}) = 50 \pm 10~\mu $eV linewidth.
Because of the intrinsic width of the mode, a splitting due to an effective single ion anisotropy, $ D $, could not be resolved, contrary to the case of the nominally pure IPA-CuCl$_3$ material, where $ D = 37~\mu $eV was observed by NRSE-TAS \cite{Nafradi2011}.
However, the mode has clearly not disappeared entirely, as was conjectured based on the absence of the low temperature electron spin resonance signal \cite{Manaka2008}.
The observed $\Gamma(0.5~\mathrm{K})=50~\mu$eV energy width of the magnon mode in the Br substituted compound is comparable to the excitation energy used in these measurements \cite{Manaka2008}.

In order to follow the temperature evolution of the relatively broad line in IPA-Cu(Cl$_{0.95}$Br$_{0.05}$)$_3$, we complemented the NRSE-TAS measurements by cold-TAS experiments at elevated temperatures, where the resolution of this instrument proved to be sufficient. 
In Fig.~\ref{fig:1} typical constant-q scans are shown as a function of temperature.
The data were fitted to a Lorentzian mode convoluted with the four-dimensional resolution function of the instrument (lines in Fig.~\ref{fig:1}).

The temperature dependence of the energy gap, $\Delta$, is shown in the top panel of Figure~\ref{fig:gap} (red symbols).
The previous results \cite{Nafradi2011,Zheludev2008} on the clean IPA-CuCl$_3$ are also shown for comparison (blue symbols).
In the Br substituted compound, $\Delta$ is always larger, but its temperature evolution is similar to the pure compound.
Figure~\ref{fig:gap}~b shows the temperature dependence of the bond randomness induced blueshift, $\delta \Delta = \Delta_{5\%~{\mathrm{Br}}}-\Delta_{0\%~{\mathrm{Br}}}$ in the same temperature range.
For $ T > 13 $~K, $\delta \Delta$ is about 0.25~meV, and it is temperature independent within our experimental precision.
Upon lowering the temperature, $ \delta \Delta $ drops by a factor of three in a narrow temperature range of $ 8 \lesssim T \lesssim 13 $~K. For $ T \lesssim 8 $~K, $\delta \Delta$ saturates at $\sim 0.1$~meV.

The temperature dependence of the magnon linewidth, $\Gamma$, extracted from the fits for IPA-Cu(Cl$_{0.95}$Br$_{0.05}$)$_3$ is also similar to IPA-CuCl$_3$ in the entire temperature range, although the resonance in the Br-substituted system is always broader (Fig.~\ref{fig:3}~a). 
The bond randomness induced magnon decoherence $ \delta \Gamma = \Gamma_{5\%~\mathrm{Br}}-\Gamma_{0\%~\mathrm{Br}}$ is plotted in Fig.~\ref{fig:3}~b as a function of temperature.
Above $ T~\sim 13 $~K, it is temperature independent ($ \delta \Gamma \sim 0.13$~meV) within the experimental error, and then decreases quickly upon cooling between $ 8 \lesssim T \lesssim 13 $~K.
A temperature independent value $ \delta \Gamma \sim 0.05$~meV is again observed for $ T \lesssim 8 $~K.
Notably $\delta \Delta \propto \delta \Gamma$ in the entire temperature range, in accordance with the behavior of the pure system.
%

%

We discuss these observations in the framework of the dimensional crossover picture developed on the basis of prior TAS and NRSE-TAS data on pure IPA-CuCl$_3$, which are well described in the entire temperature range by a mean-field analysis of a spin-1 chain with weak inter-chain interactions on a cubic lattice \cite{Nafradi2011,Senechal1993a,Senechal1993}. 
(Note that the predictions of the NL$\sigma$M hold for both spin-1 chains and spin-1/2 ladders.) 
The results of this analysis are shown in Fig.~\ref{fig:gap}~a) by a blue line. 
For temperatures exceeding the energy scale of the inter-ladder interactions, $J_4/k_B $, the magnons follow an effective 1D dispersion relation, and the temperature dependence of $\Delta$ and $\Gamma$ is in good agreement with the predictions of the 1D NL$\sigma$M. For $T << J_4/k_B $, on the other hand, the 2D dispersion is fully established, and thermally excited magnons are confined to its 2D troughs. A crossover from 1D to 2D behavior takes place at intermediate temperatures.

The observation of approximately $T$-independent values of $\delta \Delta$ and $\delta \Gamma$ in the Br-substituted system at high temperatures is qualitatively consistent with this picture, and with Matthiessen's rule which has been extensively applied in the analysis of magnon-mediated heat conduction in 1D and 2D spin systems \cite{Hess_PRL_2004,Hess_PRB_2006}. 
Written in terms of the magnon mean free path, $L$, this rule stipulates that $ L^{-1} = L_{\mathrm{dynamic}}^{-1} + L_{\mathrm{static}}^{-1}$, where $L_{\mathrm{dynamic}}$ and $L_{\mathrm{static}}$ arise from magnon-magnon and magnon-defect collisions, respectively. 
At high temperatures, the $T$-dependence of $L$ arises solely from $L_{\mathrm{dynamic}}$, which is inversely proportional to the density of thermally excited magnons and is well described by the NL$\sigma$M. 
We can thus compare our experimental data to the results of Monte Carlo calculations for finite spin-1 chains \cite{Nightingale1986}, where the observed value of $\delta \Delta / \Delta_0 \sim 0.2$ corresponds to a chain length of $L_{\mathrm{static}} \sim 10 a$. 
A similar estimate can be made on the basis of the broadening $ \delta \Gamma / \Delta_0 \sim 0.1$ observed at high temperature, which together with the magnon velocity, $ v = 2.9 $~meV,  \cite{Zheludev2008} translates into $L_{\mathrm{static}} \sim 22 a$. 
The difference between these two estimates of $L_{\mathrm{static}}$ may be influenced by the unresolved splitting of the magnon resonance in IPA-Cu(Cl$_{1-x}$Br$_x$)$_3$. 
Both values are larger than $5 a$, the average distance between Br defects along the ladders in our sample. 
The deviations between the theoretical predictions for finite-size systems and the experimental observations on IPA-Cu(Cl$_{0.95}$Br$_{0.05}$)$_3$ indicate that not all Br defects effectively break up the chains into finite-size segments. 
In particular, defects on the legs of the ladder may be less effective scattering centers than those along the rungs and diagonals.

At temperatures 8 K $\lesssim T \lesssim 13$ K, $\delta \Delta$ and $\delta \Gamma$ are strongly $T$-dependent, which indicates that the simple Matthiessen rule no longer applies in this regime. 
This is consistent with the dimensional crossover picture, according to which the effective dimensionality of the spin system, and hence also the character of the scattering of magnons from bond defects, is temperature dependent in this temperature range. 
This also implies that thermal conductivity data should be interpreted with caution in quasi-1D magnets, where dimensional crossover phenomena are common.

In the low-$T$ limit, the density of magnons thermally excited across the gap decreases exponentially upon cooling, and $L$ is expected to be dominated by the static contribution, $L_{\mathrm{static}}$. 
The observation of temperature-dependent values of  $\delta \Delta$ and $\delta \Gamma$ for $T \lesssim 8$ K is consistent with this expectation. 
The reduced values of $\delta \Delta$ and $\delta \Gamma$ at low temperatures, compared to those observed in the high-$T$ regime, are again in qualitative agreement with the dimensional crossover picture, and with the notion that defects are less disruptive for 2D than for 1D spin correlations. 
A quantitative description of these data will require calculations based on the specific spin Hamiltonian for IPA-CuCl$_3$. 
We hope that our data will stimulate further experimental and theoretical work on dimensional crossover phenomena and their influence on the magnon spectrum and magnon-mediated thermal transport in IPA-CuCl$_3$ and other quasi-1D quantum magnets.

\begin{acknowledgments}
We thank K. Buchner for technical assistance and the German DFG for financial support under grant No. SFB/TRR 80.
N.B. acknowledges support from the Prospective Research program No. PBELP2-125427 of the Swiss NSF.
H.M. is supported by a Grant-to-Aid for Young Scientist JSPS.
\end{acknowledgments}


\begin{thebibliography}{26}%
\makeatletter
\providecommand \@ifxundefined [1]{%
 \@ifx{#1\undefined}
}%
\providecommand \@ifnum [1]{%
 \ifnum #1\expandafter \@firstoftwo
 \else \expandafter \@secondoftwo
 \fi
}%
\providecommand \@ifx [1]{%
 \ifx #1\expandafter \@firstoftwo
 \else \expandafter \@secondoftwo
 \fi
}%
\providecommand \natexlab [1]{#1}%
\providecommand \enquote  [1]{``#1''}%
\providecommand \bibnamefont  [1]{#1}%
\providecommand \bibfnamefont [1]{#1}%
\providecommand \citenamefont [1]{#1}%
\providecommand \href@noop [0]{\@secondoftwo}%
\providecommand \href [0]{\begingroup \@sanitize@url \@href}%
\providecommand \@href[1]{\@@startlink{#1}\@@href}%
\providecommand \@@href[1]{\endgroup#1\@@endlink}%
\providecommand \@sanitize@url [0]{\catcode `\\12\catcode `\$12\catcode
  `\&12\catcode `\#12\catcode `\^12\catcode `\_12\catcode `\%12\relax}%
\providecommand \@@startlink[1]{}%
\providecommand \@@endlink[0]{}%
\providecommand \url  [0]{\begingroup\@sanitize@url \@url }%
\providecommand \@url [1]{\endgroup\@href {#1}{\urlprefix }}%
\providecommand \urlprefix  [0]{URL }%
\providecommand \Eprint [0]{\href }%
\providecommand \doibase [0]{http://dx.doi.org/}%
\providecommand \selectlanguage [0]{\@gobble}%
\providecommand \bibinfo  [0]{\@secondoftwo}%
\providecommand \bibfield  [0]{\@secondoftwo}%
\providecommand \translation [1]{[#1]}%
\providecommand \BibitemOpen [0]{}%
\providecommand \bibitemStop [0]{}%
\providecommand \bibitemNoStop [0]{.\EOS\space}%
\providecommand \EOS [0]{\spacefactor3000\relax}%
\providecommand \BibitemShut  [1]{\csname bibitem#1\endcsname}%
\let\auto@bib@innerbib\@empty
\bibitem [{\citenamefont {Altshuler}\ \emph {et~al.}(2006)\citenamefont
  {Altshuler}, \citenamefont {Konik},\ and\ \citenamefont
  {Tsvelik}}]{Altshuler2006}%
  \BibitemOpen
  \bibfield  {author} {\bibinfo {author} {\bibfnamefont {B.}~\bibnamefont
  {Altshuler}}, \bibinfo {author} {\bibfnamefont {R.}~\bibnamefont {Konik}}, \
  and\ \bibinfo {author} {\bibfnamefont {A.}~\bibnamefont {Tsvelik}},\ }\href
  {\doibase DOI: 10.1016/j.nuclphysb.2006.01.022} {\bibfield  {journal}
  {\bibinfo  {journal} {Nuclear Physics B}\ }\textbf {\bibinfo {volume}
  {739}},\ \bibinfo {pages} {311 } (\bibinfo {year} {2006})}\BibitemShut
  {NoStop}%
\bibitem [{\citenamefont {Hess}\ \emph {et~al.}(2001)\citenamefont {Hess},
  \citenamefont {Baumann}, \citenamefont {Ammerahl}, \citenamefont {B\"uchner},
  \citenamefont {Heidrich-Meisner}, \citenamefont {Brenig},\ and\ \citenamefont
  {Revcolevschi}}]{Hess_PRB_2001}%
  \BibitemOpen
  \bibfield  {author} {\bibinfo {author} {\bibfnamefont {C.}~\bibnamefont
  {Hess}}, \bibinfo {author} {\bibfnamefont {C.}~\bibnamefont {Baumann}},
  \bibinfo {author} {\bibfnamefont {U.}~\bibnamefont {Ammerahl}}, \bibinfo
  {author} {\bibfnamefont {B.}~\bibnamefont {B\"uchner}}, \bibinfo {author}
  {\bibfnamefont {F.}~\bibnamefont {Heidrich-Meisner}}, \bibinfo {author}
  {\bibfnamefont {W.}~\bibnamefont {Brenig}}, \ and\ \bibinfo {author}
  {\bibfnamefont {A.}~\bibnamefont {Revcolevschi}},\ }\href {\doibase
  10.1103/PhysRevB.64.184305} {\bibfield  {journal} {\bibinfo  {journal} {Phys.
  Rev. B}\ }\textbf {\bibinfo {volume} {64}},\ \bibinfo {pages} {184305}
  (\bibinfo {year} {2001})}\BibitemShut {NoStop}%
\bibitem [{\citenamefont {Senechal}(1993{\natexlab{a}})}]{Senechal1993}%
  \BibitemOpen
  \bibfield  {author} {\bibinfo {author} {\bibfnamefont {D.}~\bibnamefont
  {Senechal}},\ }\href@noop {} {\bibfield  {journal} {\bibinfo  {journal}
  {Phys. Rev. B}\ }\textbf {\bibinfo {volume} {48}},\ \bibinfo {pages} {15880}
  (\bibinfo {year} {1993}{\natexlab{a}})}\BibitemShut {NoStop}%
\bibitem [{\citenamefont {Sachdev}\ and\ \citenamefont
  {Damle}(1997)}]{Sachdev1997}%
  \BibitemOpen
  \bibfield  {author} {\bibinfo {author} {\bibfnamefont {S.}~\bibnamefont
  {Sachdev}}\ and\ \bibinfo {author} {\bibfnamefont {K.}~\bibnamefont
  {Damle}},\ }\href@noop {} {\bibfield  {journal} {\bibinfo  {journal} {Phys.
  Rev. Lett.}\ }\textbf {\bibinfo {volume} {78}},\ \bibinfo {pages} {943}
  (\bibinfo {year} {1997})}\BibitemShut {NoStop}%
\bibitem [{\citenamefont {Zheludev}\ \emph {et~al.}(2008)\citenamefont
  {Zheludev}, \citenamefont {Garlea}, \citenamefont {Regnault}, \citenamefont
  {Manaka}, \citenamefont {Tsvelik},\ and\ \citenamefont
  {Chung}}]{Zheludev2008}%
  \BibitemOpen
  \bibfield  {author} {\bibinfo {author} {\bibfnamefont {A.}~\bibnamefont
  {Zheludev}}, \bibinfo {author} {\bibfnamefont {V.~O.}\ \bibnamefont
  {Garlea}}, \bibinfo {author} {\bibfnamefont {L.~P.}\ \bibnamefont
  {Regnault}}, \bibinfo {author} {\bibfnamefont {H.}~\bibnamefont {Manaka}},
  \bibinfo {author} {\bibfnamefont {A.}~\bibnamefont {Tsvelik}}, \ and\
  \bibinfo {author} {\bibfnamefont {J.~H.}\ \bibnamefont {Chung}},\ }\href@noop
  {} {\bibfield  {journal} {\bibinfo  {journal} {Phys. Rev. Lett.}\ }\textbf
  {\bibinfo {volume} {100}},\ \bibinfo {pages} {157204} (\bibinfo {year}
  {2008})}\BibitemShut {NoStop}%
\bibitem [{\citenamefont {Xu}\ \emph {et~al.}(2007)\citenamefont {Xu},
  \citenamefont {Broholm}, \citenamefont {Soh}, \citenamefont {Aeppli},
  \citenamefont {DiTusa}, \citenamefont {Chen}, \citenamefont {Kenzelmann},
  \citenamefont {Frost}, \citenamefont {Ito}, \citenamefont {Oka},\ and\
  \citenamefont {Takagi}}]{Xu2007}%
  \BibitemOpen
  \bibfield  {author} {\bibinfo {author} {\bibfnamefont {G.~Y.}\ \bibnamefont
  {Xu}}, \bibinfo {author} {\bibfnamefont {C.}~\bibnamefont {Broholm}},
  \bibinfo {author} {\bibfnamefont {Y.~A.}\ \bibnamefont {Soh}}, \bibinfo
  {author} {\bibfnamefont {G.}~\bibnamefont {Aeppli}}, \bibinfo {author}
  {\bibfnamefont {J.~F.}\ \bibnamefont {DiTusa}}, \bibinfo {author}
  {\bibfnamefont {Y.}~\bibnamefont {Chen}}, \bibinfo {author} {\bibfnamefont
  {M.}~\bibnamefont {Kenzelmann}}, \bibinfo {author} {\bibfnamefont {C.~D.}\
  \bibnamefont {Frost}}, \bibinfo {author} {\bibfnamefont {T.}~\bibnamefont
  {Ito}}, \bibinfo {author} {\bibfnamefont {K.}~\bibnamefont {Oka}}, \ and\
  \bibinfo {author} {\bibfnamefont {H.}~\bibnamefont {Takagi}},\ }\href@noop {}
  {\bibfield  {journal} {\bibinfo  {journal} {Science}\ }\textbf {\bibinfo
  {volume} {317}},\ \bibinfo {pages} {1049} (\bibinfo {year}
  {2007})}\BibitemShut {NoStop}%
\bibitem [{\citenamefont {N\'afr\'adi}\ \emph {et~al.}(2011)\citenamefont
  {N\'afr\'adi}, \citenamefont {Keller}, \citenamefont {Manaka}, \citenamefont
  {Zheludev},\ and\ \citenamefont {Keimer}}]{Nafradi2011}%
  \BibitemOpen
  \bibfield  {author} {\bibinfo {author} {\bibfnamefont {B.}~\bibnamefont
  {N\'afr\'adi}}, \bibinfo {author} {\bibfnamefont {T.}~\bibnamefont {Keller}},
  \bibinfo {author} {\bibfnamefont {H.}~\bibnamefont {Manaka}}, \bibinfo
  {author} {\bibfnamefont {A.}~\bibnamefont {Zheludev}}, \ and\ \bibinfo
  {author} {\bibfnamefont {B.}~\bibnamefont {Keimer}},\ }\href {\doibase
  10.1103/PhysRevLett.106.177202} {\bibfield  {journal} {\bibinfo  {journal}
  {Phys. Rev. Lett.}\ }\textbf {\bibinfo {volume} {106}},\ \bibinfo {pages}
  {177202} (\bibinfo {year} {2011})}\BibitemShut {NoStop}%
\bibitem [{\citenamefont {Hong}\ \emph
  {et~al.}(2010{\natexlab{a}})\citenamefont {Hong}, \citenamefont {Zheludev},
  \citenamefont {Manaka},\ and\ \citenamefont {Regnault}}]{Hong2010}%
  \BibitemOpen
  \bibfield  {author} {\bibinfo {author} {\bibfnamefont {T.}~\bibnamefont
  {Hong}}, \bibinfo {author} {\bibfnamefont {A.}~\bibnamefont {Zheludev}},
  \bibinfo {author} {\bibfnamefont {H.}~\bibnamefont {Manaka}}, \ and\ \bibinfo
  {author} {\bibfnamefont {L.-P.}\ \bibnamefont {Regnault}},\ }\href@noop {}
  {\bibfield  {journal} {\bibinfo  {journal} {Phys. Rev. B}\ }\textbf {\bibinfo
  {volume} {81}},\ \bibinfo {pages} {060410(R)} (\bibinfo {year}
  {2010}{\natexlab{a}})}\BibitemShut {NoStop}%
\bibitem [{\citenamefont {Cr\'epin}\ \emph {et~al.}(2011)\citenamefont
  {Cr\'epin}, \citenamefont {Laflorencie}, \citenamefont {Roux},\ and\
  \citenamefont {Simon}}]{Crepin2011}%
  \BibitemOpen
  \bibfield  {author} {\bibinfo {author} {\bibfnamefont {F.}~\bibnamefont
  {Cr\'epin}}, \bibinfo {author} {\bibfnamefont {N.}~\bibnamefont
  {Laflorencie}}, \bibinfo {author} {\bibfnamefont {G.}~\bibnamefont {Roux}}, \
  and\ \bibinfo {author} {\bibfnamefont {P.}~\bibnamefont {Simon}},\ }\href
  {\doibase 10.1103/PhysRevB.84.054517} {\bibfield  {journal} {\bibinfo
  {journal} {Phys. Rev. B}\ }\textbf {\bibinfo {volume} {84}},\ \bibinfo
  {pages} {054517} (\bibinfo {year} {2011})}\BibitemShut {NoStop}%
\bibitem [{\citenamefont {Carrasquilla}\ \emph {et~al.}(2011)\citenamefont
  {Carrasquilla}, \citenamefont {Becca},\ and\ \citenamefont
  {Fabrizio}}]{Carrasquilla2011}%
  \BibitemOpen
  \bibfield  {author} {\bibinfo {author} {\bibfnamefont {J.}~\bibnamefont
  {Carrasquilla}}, \bibinfo {author} {\bibfnamefont {F.}~\bibnamefont {Becca}},
  \ and\ \bibinfo {author} {\bibfnamefont {M.}~\bibnamefont {Fabrizio}},\
  }\href {\doibase 10.1103/PhysRevB.83.245101} {\bibfield  {journal} {\bibinfo
  {journal} {Phys. Rev. B}\ }\textbf {\bibinfo {volume} {83}},\ \bibinfo
  {pages} {245101} (\bibinfo {year} {2011})}\BibitemShut {NoStop}%
\bibitem [{\citenamefont {Hong}\ \emph
  {et~al.}(2010{\natexlab{b}})\citenamefont {Hong}, \citenamefont {Kim},
  \citenamefont {Hotta}, \citenamefont {Takano}, \citenamefont {Tremelling},
  \citenamefont {Turnbull}, \citenamefont {Landee}, \citenamefont {Kang},
  \citenamefont {Christensen}, \citenamefont {Lefmann}, \citenamefont
  {Schmidt}, \citenamefont {Uhrig},\ and\ \citenamefont {Broholm}}]{Hong2010b}%
  \BibitemOpen
  \bibfield  {author} {\bibinfo {author} {\bibfnamefont {T.}~\bibnamefont
  {Hong}}, \bibinfo {author} {\bibfnamefont {Y.~H.}\ \bibnamefont {Kim}},
  \bibinfo {author} {\bibfnamefont {C.}~\bibnamefont {Hotta}}, \bibinfo
  {author} {\bibfnamefont {Y.}~\bibnamefont {Takano}}, \bibinfo {author}
  {\bibfnamefont {G.}~\bibnamefont {Tremelling}}, \bibinfo {author}
  {\bibfnamefont {M.~M.}\ \bibnamefont {Turnbull}}, \bibinfo {author}
  {\bibfnamefont {C.~P.}\ \bibnamefont {Landee}}, \bibinfo {author}
  {\bibfnamefont {H.-J.}\ \bibnamefont {Kang}}, \bibinfo {author}
  {\bibfnamefont {N.~B.}\ \bibnamefont {Christensen}}, \bibinfo {author}
  {\bibfnamefont {K.}~\bibnamefont {Lefmann}}, \bibinfo {author} {\bibfnamefont
  {K.~P.}\ \bibnamefont {Schmidt}}, \bibinfo {author} {\bibfnamefont {G.~S.}\
  \bibnamefont {Uhrig}}, \ and\ \bibinfo {author} {\bibfnamefont
  {C.}~\bibnamefont {Broholm}},\ }\href {\doibase
  10.1103/PhysRevLett.105.137207} {\bibfield  {journal} {\bibinfo  {journal}
  {Phys. Rev. Lett.}\ }\textbf {\bibinfo {volume} {105}},\ \bibinfo {pages}
  {137207} (\bibinfo {year} {2010}{\natexlab{b}})}\BibitemShut {NoStop}%
\bibitem [{\citenamefont {Trinh}\ \emph {et~al.}(2012)\citenamefont {Trinh},
  \citenamefont {Haas}, \citenamefont {Yu},\ and\ \citenamefont
  {Roscilde}}]{Trinh_PRB_2012}%
  \BibitemOpen
  \bibfield  {author} {\bibinfo {author} {\bibfnamefont {K.}~\bibnamefont
  {Trinh}}, \bibinfo {author} {\bibfnamefont {S.}~\bibnamefont {Haas}},
  \bibinfo {author} {\bibfnamefont {R.}~\bibnamefont {Yu}}, \ and\ \bibinfo
  {author} {\bibfnamefont {T.}~\bibnamefont {Roscilde}},\ }\href {\doibase
  10.1103/PhysRevB.85.035134} {\bibfield  {journal} {\bibinfo  {journal} {Phys.
  Rev. B}\ }\textbf {\bibinfo {volume} {85}},\ \bibinfo {pages} {035134}
  (\bibinfo {year} {2012})}\BibitemShut {NoStop}%
\bibitem [{\citenamefont {Greven}\ and\ \citenamefont
  {Birgeneau}(1998)}]{Greven_PRL_1998}%
  \BibitemOpen
  \bibfield  {author} {\bibinfo {author} {\bibfnamefont {M.}~\bibnamefont
  {Greven}}\ and\ \bibinfo {author} {\bibfnamefont {R.~J.}\ \bibnamefont
  {Birgeneau}},\ }\href {\doibase 10.1103/PhysRevLett.81.1945} {\bibfield
  {journal} {\bibinfo  {journal} {Phys. Rev. Lett.}\ }\textbf {\bibinfo
  {volume} {81}},\ \bibinfo {pages} {1945} (\bibinfo {year}
  {1998})}\BibitemShut {NoStop}%
\bibitem [{\citenamefont {Hess}\ \emph {et~al.}(2004)\citenamefont {Hess},
  \citenamefont {ElHaes}, \citenamefont {B\"uchner}, \citenamefont {Ammerahl},
  \citenamefont {H\"ucker},\ and\ \citenamefont
  {Revcolevschi}}]{Hess_PRL_2004}%
  \BibitemOpen
  \bibfield  {author} {\bibinfo {author} {\bibfnamefont {C.}~\bibnamefont
  {Hess}}, \bibinfo {author} {\bibfnamefont {H.}~\bibnamefont {ElHaes}},
  \bibinfo {author} {\bibfnamefont {B.}~\bibnamefont {B\"uchner}}, \bibinfo
  {author} {\bibfnamefont {U.}~\bibnamefont {Ammerahl}}, \bibinfo {author}
  {\bibfnamefont {M.}~\bibnamefont {H\"ucker}}, \ and\ \bibinfo {author}
  {\bibfnamefont {A.}~\bibnamefont {Revcolevschi}},\ }\href {\doibase
  10.1103/PhysRevLett.93.027005} {\bibfield  {journal} {\bibinfo  {journal}
  {Phys. Rev. Lett.}\ }\textbf {\bibinfo {volume} {93}},\ \bibinfo {pages}
  {027005} (\bibinfo {year} {2004})}\BibitemShut {NoStop}%
\bibitem [{\citenamefont {Hess}\ \emph {et~al.}(2006)\citenamefont {Hess},
  \citenamefont {Ribeiro}, \citenamefont {B\"uchner}, \citenamefont {ElHaes},
  \citenamefont {Roth}, \citenamefont {Ammerahl},\ and\ \citenamefont
  {Revcolevschi}}]{Hess_PRB_2006}%
  \BibitemOpen
  \bibfield  {author} {\bibinfo {author} {\bibfnamefont {C.}~\bibnamefont
  {Hess}}, \bibinfo {author} {\bibfnamefont {P.}~\bibnamefont {Ribeiro}},
  \bibinfo {author} {\bibfnamefont {B.}~\bibnamefont {B\"uchner}}, \bibinfo
  {author} {\bibfnamefont {H.}~\bibnamefont {ElHaes}}, \bibinfo {author}
  {\bibfnamefont {G.}~\bibnamefont {Roth}}, \bibinfo {author} {\bibfnamefont
  {U.}~\bibnamefont {Ammerahl}}, \ and\ \bibinfo {author} {\bibfnamefont
  {A.}~\bibnamefont {Revcolevschi}},\ }\href {\doibase
  10.1103/PhysRevB.73.104407} {\bibfield  {journal} {\bibinfo  {journal} {Phys.
  Rev. B}\ }\textbf {\bibinfo {volume} {73}},\ \bibinfo {pages} {104407}
  (\bibinfo {year} {2006})}\BibitemShut {NoStop}%
\bibitem [{\citenamefont {Manaka}\ \emph {et~al.}(2001)\citenamefont {Manaka},
  \citenamefont {Yamada},\ and\ \citenamefont {Katori}}]{Manaka2001e}%
  \BibitemOpen
  \bibfield  {author} {\bibinfo {author} {\bibfnamefont {H.}~\bibnamefont
  {Manaka}}, \bibinfo {author} {\bibfnamefont {I.}~\bibnamefont {Yamada}}, \
  and\ \bibinfo {author} {\bibfnamefont {H.~A.}\ \bibnamefont {Katori}},\
  }\href@noop {} {\bibfield  {journal} {\bibinfo  {journal} {Phys. Rev. B}\
  }\textbf {\bibinfo {volume} {63}},\ \bibinfo {pages} {104408} (\bibinfo
  {year} {2001})}\BibitemShut {NoStop}%
\bibitem [{\citenamefont {Manaka}\ \emph {et~al.}(1997)\citenamefont {Manaka},
  \citenamefont {Yamada},\ and\ \citenamefont {Yamaguchi}}]{Manaka1997c}%
  \BibitemOpen
  \bibfield  {author} {\bibinfo {author} {\bibfnamefont {H.}~\bibnamefont
  {Manaka}}, \bibinfo {author} {\bibfnamefont {I.}~\bibnamefont {Yamada}}, \
  and\ \bibinfo {author} {\bibfnamefont {K.}~\bibnamefont {Yamaguchi}},\
  }\href@noop {} {\bibfield  {journal} {\bibinfo  {journal} {J. Phys. Soc.
  Jpn.}\ }\textbf {\bibinfo {volume} {66}},\ \bibinfo {pages} {564} (\bibinfo
  {year} {1997})}\BibitemShut {NoStop}%
\bibitem [{Note1()}]{Note1}%
  \BibitemOpen
  \bibinfo {note} {Here IPA stands for isopropyl ammonium, which in our case
  was fully deuterated: (CD$_3$)$_2$CDND$_3$.}\BibitemShut {Stop}%
\bibitem [{\citenamefont {Masuda}\ \emph {et~al.}(2006)\citenamefont {Masuda},
  \citenamefont {Zheludev}, \citenamefont {Manaka}, \citenamefont {Regnault},
  \citenamefont {Chung},\ and\ \citenamefont {Qiu}}]{Masuda2006}%
  \BibitemOpen
  \bibfield  {author} {\bibinfo {author} {\bibfnamefont {T.}~\bibnamefont
  {Masuda}}, \bibinfo {author} {\bibfnamefont {A.}~\bibnamefont {Zheludev}},
  \bibinfo {author} {\bibfnamefont {H.}~\bibnamefont {Manaka}}, \bibinfo
  {author} {\bibfnamefont {L.~P.}\ \bibnamefont {Regnault}}, \bibinfo {author}
  {\bibfnamefont {J.~H.}\ \bibnamefont {Chung}}, \ and\ \bibinfo {author}
  {\bibfnamefont {Y.}~\bibnamefont {Qiu}},\ }\href@noop {} {\bibfield
  {journal} {\bibinfo  {journal} {Phys. Rev. Lett.}\ }\textbf {\bibinfo
  {volume} {96}},\ \bibinfo {pages} {047210} (\bibinfo {year}
  {2006})}\BibitemShut {NoStop}%
\bibitem [{\citenamefont {Hida}(2003)}]{Hida2003}%
  \BibitemOpen
  \bibfield  {author} {\bibinfo {author} {\bibfnamefont {K.}~\bibnamefont
  {Hida}},\ }\href@noop {} {\bibfield  {journal} {\bibinfo  {journal} {J. Phys.
  Soc. Jpn.}\ }\textbf {\bibinfo {volume} {72}},\ \bibinfo {pages} {688}
  (\bibinfo {year} {2003})}\BibitemShut {NoStop}%
\bibitem [{\citenamefont {Manaka}\ \emph {et~al.}(2008)\citenamefont {Manaka},
  \citenamefont {Kolomiets},\ and\ \citenamefont {Goto}}]{Manaka2008}%
  \BibitemOpen
  \bibfield  {author} {\bibinfo {author} {\bibfnamefont {H.}~\bibnamefont
  {Manaka}}, \bibinfo {author} {\bibfnamefont {A.~V.}\ \bibnamefont
  {Kolomiets}}, \ and\ \bibinfo {author} {\bibfnamefont {T.}~\bibnamefont
  {Goto}},\ }\href {\doibase 10.1103/PhysRevLett.101.077204} {\bibfield
  {journal} {\bibinfo  {journal} {Phys. Rev. Lett.}\ }\textbf {\bibinfo
  {volume} {101}},\ \bibinfo {pages} {077204} (\bibinfo {year}
  {2008})}\BibitemShut {NoStop}%
\bibitem [{\citenamefont {Fischer}\ \emph {et~al.}(2011)\citenamefont
  {Fischer}, \citenamefont {Duffe},\ and\ \citenamefont {Uhrig}}]{Fischer2011}%
  \BibitemOpen
  \bibfield  {author} {\bibinfo {author} {\bibfnamefont {T.}~\bibnamefont
  {Fischer}}, \bibinfo {author} {\bibfnamefont {S.}~\bibnamefont {Duffe}}, \
  and\ \bibinfo {author} {\bibfnamefont {G.~S.}\ \bibnamefont {Uhrig}},\ }\href
  {http://stacks.iop.org/0295-5075/96/i=4/a=47001} {\bibfield  {journal}
  {\bibinfo  {journal} {EPL (Europhysics Letters)}\ }\textbf {\bibinfo {volume}
  {96}},\ \bibinfo {pages} {47001} (\bibinfo {year} {2011})}\BibitemShut
  {NoStop}%
\bibitem [{\citenamefont {Keller}\ \emph {et~al.}(2002)\citenamefont {Keller},
  \citenamefont {Habicht}, \citenamefont {Klann}, \citenamefont {Schneider},\
  and\ \citenamefont {Keimer}}]{Keller2002}%
  \BibitemOpen
  \bibfield  {author} {\bibinfo {author} {\bibfnamefont {T.}~\bibnamefont
  {Keller}}, \bibinfo {author} {\bibfnamefont {K.}~\bibnamefont {Habicht}},
  \bibinfo {author} {\bibfnamefont {H.}~\bibnamefont {Klann}}, \bibinfo
  {author} {\bibfnamefont {H.}~\bibnamefont {Schneider}}, \ and\ \bibinfo
  {author} {\bibfnamefont {B.}~\bibnamefont {Keimer}},\ }\href@noop {}
  {\bibfield  {journal} {\bibinfo  {journal} {Appl. Phys. A}\ }\textbf
  {\bibinfo {volume} {74}},\ \bibinfo {pages} {s332} (\bibinfo {year}
  {2002})}\BibitemShut {NoStop}%
\bibitem [{\citenamefont {Senechal}(1993{\natexlab{b}})}]{Senechal1993a}%
  \BibitemOpen
  \bibfield  {author} {\bibinfo {author} {\bibfnamefont {D.}~\bibnamefont
  {Senechal}},\ }\href@noop {} {\bibfield  {journal} {\bibinfo  {journal}
  {Phys. Rev. B}\ }\textbf {\bibinfo {volume} {47}},\ \bibinfo {pages} {8353}
  (\bibinfo {year} {1993}{\natexlab{b}})}\BibitemShut {NoStop}%
\bibitem [{\citenamefont {Damle}\ and\ \citenamefont
  {Sachdev}(1998)}]{Damle1998}%
  \BibitemOpen
  \bibfield  {author} {\bibinfo {author} {\bibfnamefont {K.}~\bibnamefont
  {Damle}}\ and\ \bibinfo {author} {\bibfnamefont {S.}~\bibnamefont
  {Sachdev}},\ }\href@noop {} {\bibfield  {journal} {\bibinfo  {journal} {Phys.
  Rev. B}\ }\textbf {\bibinfo {volume} {57}},\ \bibinfo {pages} {8307}
  (\bibinfo {year} {1998})}\BibitemShut {NoStop}%
\bibitem [{\citenamefont {Nightingale}\ and\ \citenamefont
  {Bl\"ote}(1986)}]{Nightingale1986}%
  \BibitemOpen
  \bibfield  {author} {\bibinfo {author} {\bibfnamefont {M.~P.}\ \bibnamefont
  {Nightingale}}\ and\ \bibinfo {author} {\bibfnamefont {H.~W.~J.}\
  \bibnamefont {Bl\"ote}},\ }\href {\doibase 10.1103/PhysRevB.33.659}
  {\bibfield  {journal} {\bibinfo  {journal} {Phys. Rev. B}\ }\textbf {\bibinfo
  {volume} {33}},\ \bibinfo {pages} {659} (\bibinfo {year} {1986})}\BibitemShut
  {NoStop}%
\end{thebibliography}
%

\end{document}